\documentclass[manuscript]{aastex}
\shorttitle{PIC simulation on shock tube}
\shortauthors{Matsukiyo et al.}

\begin{document}

%% LaTeX will automatically break titles if they run longer than
%% one line. However, you may use \\ to force a line break if
%% you desire.

\title{PIC simulation of a shock tube: Implications for wave transmission in the 
heliospheric boundary region}

%% Use \author, \affil, and the \and command to format
%% author and affiliation information.
%% Note that \email has replaced the old \authoremail command
%% from AASTeX v4.0. You can use \email to mark an email address
%% anywhere in the paper, not just in the front matter.
%% As in the title, use \\ to force line breaks.

\author{S. Matsukiyo}
\affil{Fuculty of Engineering Sciences, Kyushu University,\\
    Kasuga, Fukuoka, 816-8580, Japan}
\email{matsukiy@esst.kyushu-u.ac.jp}
\author{T. Noumi}
\affil{Fuculty of Engineering Sciences, Kyushu University,\\
    Kasuga, Fukuoka, 816-8580, Japan}
\email{nomi@esst.kyushu-u.ac.jp}
\author{G. P. Zank}
\affil{Center for Space Plasma and Aeronomie Research (CSPAR), University of Alabama in Huntsville,\\
    AL 35805, USA}
\email{garyp.zank@gmail.com}
\author{H. Washimi}
\affil{Center for Space Plasma and Aeronomie Research (CSPAR), University of Alabama in Huntsville,\\
    AL 35805, USA}
\email{hwashimi2004@hotmail.com}
\author{T. Hada}
\affil{Fuculty of Engineering Sciences, Kyushu University,\\
    Kasuga, Fukuoka, 816-8580, Japan}
\email{hada@esst.kyushu-u.ac.jp}

%% Mark off your abstract in the ``abstract'' environment. In the manuscript
%% style, abstract will output a Received/Accepted line after the
%% title and affiliation information. No date will appear since the author
%% does not have this information. The dates will be filled in by the
%% editorial office after submission.  Svensmark(1998)

\begin{abstract}
A shock tube problem is solved  numerically by using one-dimensional full particle-in-cell 
simulations under the condition that a relatively tenuous and weakly magnetized 
plasma is continuously pushed by a relatively dense and strongly magnetized plasma 
having supersonic relative velocity. A forward and a reverse shock and a contact 
discontinuity are self-consistently reproduced. The spatial width of the contact 
discontinuity increases as the angle between the discontinuity normal and ambient 
magnetic field decreases. The inner structure of the discontinuity shows different 
profiles between magnetic field and plasma density, or pressure, which is caused by a 
non-MHD effect of the local plasma. The region between the two shocks is turbulent. 
The fluctuations in the relatively dense plasma are compressible and propagating away 
from the contact discontinuity, although the fluctuations in the relatively tenuous 
plasma contain both compressible and incompressible components. The source of 
the compressible fluctuations in the relatively dense plasma is in the relatively tenuous 
plasma. Only compressible fast mode fluctuations generated in the relatively tenuous 
plasma are transmitted through the contact discontinuity and 
propagate in the relatively dense plasma. These fast mode fluctuations 
are steepened when passing the contact discontinuity. This wave steepening and 
probably other effects may cause the broadening of the wave spectrum 
in the very local interstellar medium plasma. The results are discussed in the context 
of the heliospheric boundary region or heliopause.
\end{abstract}

%The positively correlated fluctuations are produced in the shock front through the self-reformation process, while the negatively correlated ones are generated through mirror instability. 
%% Keywords should appear after the \end{abstract} command. The uncommented
%% example has been keyed in ApJ style. See the instructions to authors
%% for the journal to which you are submitting your paper to determine
%% what keyword punctuation is appropriate.

%% From the front matter, we move on to the body of the paper.
%% In the first two sections, notice the use of the natbib \citep
%% and \citet commands to identify citations.  The citations are
%% tied to the reference list via symbolic KEYs. The KEY corresponds
%% to the KEY in the \bibitem in the reference list below. We have
%% chosen the first three characters of the first author's name plus
%% the last two numeral of the year of publication as our KEY for
%% each reference.

%% Authors who wish to have the most important objects in their paper
%% linked in the electronic edition to a data center may do so by tagging
%% their objects with \objectname{} or \object{}.  Each macro takes the
%% object name as its required argument. The optional, square-bracket 
%% argument should be used in cases where the data center identification
%% differs from what is to be printed in the paper.  The text appearing 
%% in curly braces is what will appear in print in the published paper. 
%% If the object name is recognized by the data centers, it will be linked
%% in the electronic edition to the object data available at the data centers  
%%
%% Note that for sources with brackets in their names, e.g. [WEG2004] 14h-090,
%% the brackets must be escaped with backslashes when used in the first
%% square-bracket argument, for instance, \object[\[WEG2004\] 14h-090]{90}).
%%  Otherwise, LaTeX will issue an error. 

%%%%%%%%
\section{Introduction}
%%%%%%%%

The heliospheric boundary region plays very important roles in the transport, conversion, 
and exchange of energy and matter between the heliosphere and interstellar space. 
The region has been explored in-situ by the Voyager spacecraft. Voyager 1 crossed the 
termination shock of the solar wind in 2004 \citep{stone05, decker05, gurnet05, burlaga05} 
and heliopause in 2012 \citep{krimigis13, burlaga13, stone13, gurnet13}, and 
has since traveled in the very local interstellar medium (VLISM). Voyager 2 
crossed the termination shock in 2007 \citep{richardson08, decker08, stone08, 
burlaga08, gurnet08}, and it recently crossed the heliopause in 2018. 
The heliopause is a contact discontinuity that separates the solar wind plasma 
and the interstellar plasma from the perspective of magnetohydrodynamics 
(MHD). 

There are a number of unresolved issues regarding the structure of the boundary region 
since it appears to depart from what is expected from simple MHD theory. It is well known, 
for instance, that the behavior of the magnetic field and cosmic ray particles in the 
heliopause transition region observed by Voyager 1 is not well correlated 
\citep{burlaga14}. Since the Voyager spacecraft have explored the boundary 
region along only two radial paths, the detailed structure of the region has been 
studied by using numerical simulations. In particular kinetic scale structures 
are discussed using hybrid and full particle-in-cell (PIC) simulations. Hybrid 
simulations of the termination shock were presented by many authors 
\citep{liewer93, liewer95, kucharek95, lipatov98, lipatov99, wu09, wu10, liu10, 
giacalone10, giacalone10b} focusing on ion kinetic effects, while PIC simulations 
were used to address both ion and electron dynamics at the heliospheric 
termination shock \citep{lee05, 
matsukiyo07, matsukiyo11, oka11, matsukiyo14, yang15, lembege18}. 
Kinetic simulations of the region beyond the termination shock such as 
heliosheath, heliopause, etc., have also been undertaken, although not so extensively. 
Magnetic reconnection in the inner heliosheath is discussed using hybrid 
simulations \citep{burgess16} and PIC simulations \citep{drake17}. The stability of 
the pickup ion distribution function has been examined for various kinetic instabilities 
by using hybrid \citep{florinski10, liu12, florinski16, min18} and PIC 
\citep{niemiec16} simulations.

Because kinetic numerical simulations are very costly, they are usually used 
to reproduce a local structure. However, there are some phenomena for 
which a non-local effect is expected to be essential. \cite{burlaga15, burlaga18} showed 
that Voyager 1 observed compressible weak magnetic fluctuations in the VLISM 
immediately outside the heliopause even during a period during which there were no large 
solar wind disturbances, and they discussed the possibility that the 
fluctuations are generated in the inner heliosheath and passed across the heliopause. 
This idea is supported theoretically by \cite{zank17}. In order to reproduce a 
non-local effect in a numerical simulation, a large system size is necessary in at 
least one appropriate direction. In this study we investigate the radial structure 
of a boundary region including a shock and a contact discontinuity as in the 
heliospheric boundary region with a termination shock and heliopause, by using a 
one-dimensional PIC simulation. Kinetic properties at the boundary 
of two plasmas in contact with each other, i.e., at a contact discontinuity, are investigated 
including mixing and the radial non-local effect of wave propagation. 
To mimic the heliospheric boundary region, 
we numerically solve the so-called shock tube problem. As explained in the 
next section, it reproduces the evolution of a system that includes three 
discontinuities, a forward and a reverse shock and a contact discontinuity.

% Furthermore, kinetic simulation is useful 
%to understand inner structure of a discontinuity, like heliopause and termination 
%shock. Hence, in this study we focus on the radial kinetic structure of the 
%system including a shock and a contact discontinuity simultaneously by using 
%one-dimensional PIC simulation. To mimic the heliospheric boundary region, 
%we numerically solve the so-called shock tube problem. As explained in the 
%next section, it reproduces the evolution of a system including three 
%discontinuities, a forward and a reverse shocks and a contact discontinuity. 

The paper is organized as follows. In section 2, simulation settings and 
parameters are presented. The results of the simulations are discussed 
in section 3. Then, the summary and discussion are given in section 4.

%%%%%%%%
\section{Simulation Settings}
%%%%%%%%

The shock tube problem is solved numerically using a one-dimensional PIC code. 
Initially, the system is divided into two regions. The region with $X > X_0$ is 
filled with a relatively tenuous and weakly magnetized plasma at rest, while the 
region with $X < X_0$ is filled with a relatively dense and strongly magnetized 
plasma flowing with constant bulk velocity. The left boundary at $X=0$ is open 
and fresh dense plasma is injected continuously. At the right boundary 
($X=L$), the plasma is reflected and electromagnetic fields are absorbed. 
The distribution functions of both plasmas are assumed to be (shifted) 
Maxwellians. With the above initial conditions, three discontinuities are self-consistently 
produced as time passes propagating from left to right. The rightmost one is 
a forward shock, the leftmost one a reverse shock, and the middle one is a 
contact discontinuity, respectively. 

The simulation parameters are as follows. The plasma density from the left to the right 
is 9, the relative temperature is 4, and the relative strength of tangential magnetic 
field is 6, respectively. The ion to electron mass ratio is $m_i / m_e = 25$ in 
both plasmas. 
%${\bf 
Although this mass ratio is unrealistically small, 
electron and ion phenomena are sufficiently well separated to qualitatively resolve the 
corresponding structures. We confirmed that the structures of electromagnetic 
fields as well as particle phase space distributions are essentially unchanged 
when the mass ratio is increased to 100 for Run 3 below (not shown).
%$} 
In the left plasma, the ratio of electron plasma to cyclotron 
frequencies is 2.37 and plasma beta is 0.225, and the temperature is identical 
for both electrons and ions. While the above parameters are fixed, we perform 
five runs with different angles $\Theta_{Bn}$ of ambient magnetic field with respect to the 
$X$-axis in the $X-Z$ plane. The above relative values of density, 
temperature, and magnetic field strength are qualitatively similar to those 
between the local interstellar medium (LISM) and solar wind. Therefore, 
the left plasma mimics the local interstellar medium, while the 
right plasma mimics the solar wind plasma. In this context the forward and 
reverse shocks correspond to the solar wind termination shock and 
the bow shock in the LISM, and the contact discontinuity to the heliopause, 
respectively. Hereafter, we refer to the left plasma as interstellar (IS) plasma 
and the right plasma as solar wind (SW) plasma for convenience.

The system size is $L = 506 c/\omega_{pi,SW}$ and is divided into 80,000 
grid points. Here, $c$ is the speed of light, $\omega_{pi,SW}$ denotes the ion 
plasma frequency in the solar wind, and the grid size is a little smaller than 
the solar wind Debye length, $\Delta X = 0.63\lambda_{De,SW}$. 
We recognize that the system size is too small to compare the results with the observations quantitatively, even when limiting our discussions to essentially one-dimensional phenomena. Nonetheless, the simulations serve to illustrate what we regard as the underlying essential physical phenomena that are qualitatively important in the boundary region. 
We emphasize that the system size treated here is not large in the sense of the global heliospheric structure and therefore represents a small 1D spatial ``slice'' of the actual heliosphere.  We do not presume to address the global physics of the heliosphere in this paper, since the question of what wave fluctuations will be transmitted across the heliosheath has to be addressed on kinetic scales corresponding to these fluctuations. Accordingly, a coarsely resolved global simulation cannot be used to investigate the transmission of short wavelength fluctuations across the heliopause. For this reason, we focus on a spatially 1D simulation. 

We note also that other aspects underpinning the physics of the large-scale global heliospheric interaction with the local interstellar medium are not important to the specific problem of compressible wave generation in the VLISM. For example, the effect of collisions will not change the results discussed below. For charge-exchange collisions, the inner heliosheath, the neutral H - proton charge exchange mean free path is $\sim 1000$ AU, and in the VLISM, neutral H - proton charge exchange mean free path is $\sim 60$ AU. Therefore, for the kinetic scales of interest in this paper, charge exchange is irrelevant. Non-charge exchange collisionality of the VLISM has been recognized recently as an interesting further aspect of the physics \citep{mostafavi18}. However, the collisional length scale of proton-proton collisions, for example, is ~0.3 au, meaning that on rather larger scales, the VLISM can be regarded as collisional. On the smaller scales of interest here, by contrast, we may regard the VLISM as collisionless.

In the interstellar medium, including the VLISM, galactic cosmic rays are important and contribute to the total energy balance of the plasma. However, galactic cosmic rays are largely excluded from the inner heliosheath \citep{florinski03, burlaga13} and can safely be neglected in the simulations below. In the VLISM in the neighborhood of the heliopause, on kinetic scales, even low energy cosmic rays are completely decoupled from the thermal interstellar plasma - only on scales associated with the diffusive length scale (typically given by the spatial diffusion coefficient $\kappa$ divided by the characteristic flow speed $U$ i.e., $\kappa /U \sim 10^{24} [\mbox{m}^2/\mbox{s}] /20 [\mbox{km/s}] = 5 \times 10^{20}$ m) do cosmic rays couple dynamically to the background plasma. Consequently, since this scale length far exceeds the fluctuation scales of interest, cosmic rays are irrelevant to the problem at hand. 
Physical parameters used in the simulation are summarized in Table \ref{table1}.

%
% tabel 1
\begin{table}
\caption{\label{table1}Simulation Parameters}
\begin{tabular}{cccccc}
\hline
\hline
Run & 1 & 2 & 3 & 4 & 5 \\
\hline
$\Theta_{Bn,IS}$ & 90.0 & 87.5 & 85.0 & 82.5 & 80.0\\
$\Theta_{Bn,SW}$ & 90.0 & 75.3 & 62.3 & 51.7 & 43.4\\
$\beta_{IS}$ & 0.225 & 0.225 & 0.223 & 0.221 & 0.218\\
$\beta_{SW}$ & 0.225 & 0.211 & 0.178 & 0.141 & 0.109\\
$\omega_{pe,IS}/\Omega_{e,IS}$ & 2.37 & 2.37 & 2.36 & 2.35 & 2.34\\
$\omega_{pe,SW}/\Omega_{e,SW}$ & 4.74 & 4.59 & 4.22 & 3.75 & 3.31\\
$M_{A,BS}$ & 1.2 & 1.2 & 1.2 & 1.2 & 1.2\\
$M_{A,TS}$ & 6.5 & 6.1 & 5.6 & 4.5 & 4.0\\
\hline
\hline
\end{tabular}
\end{table}
%

%%%%%%%%%%%
\section{Simulation Results}
%%%%%%%%%%%
%%%%%%%%%%%
\subsection{Overview (Run 1)}
%%%%%%%%%%%

\begin{figure}
\epsscale{0.5}
\plotone{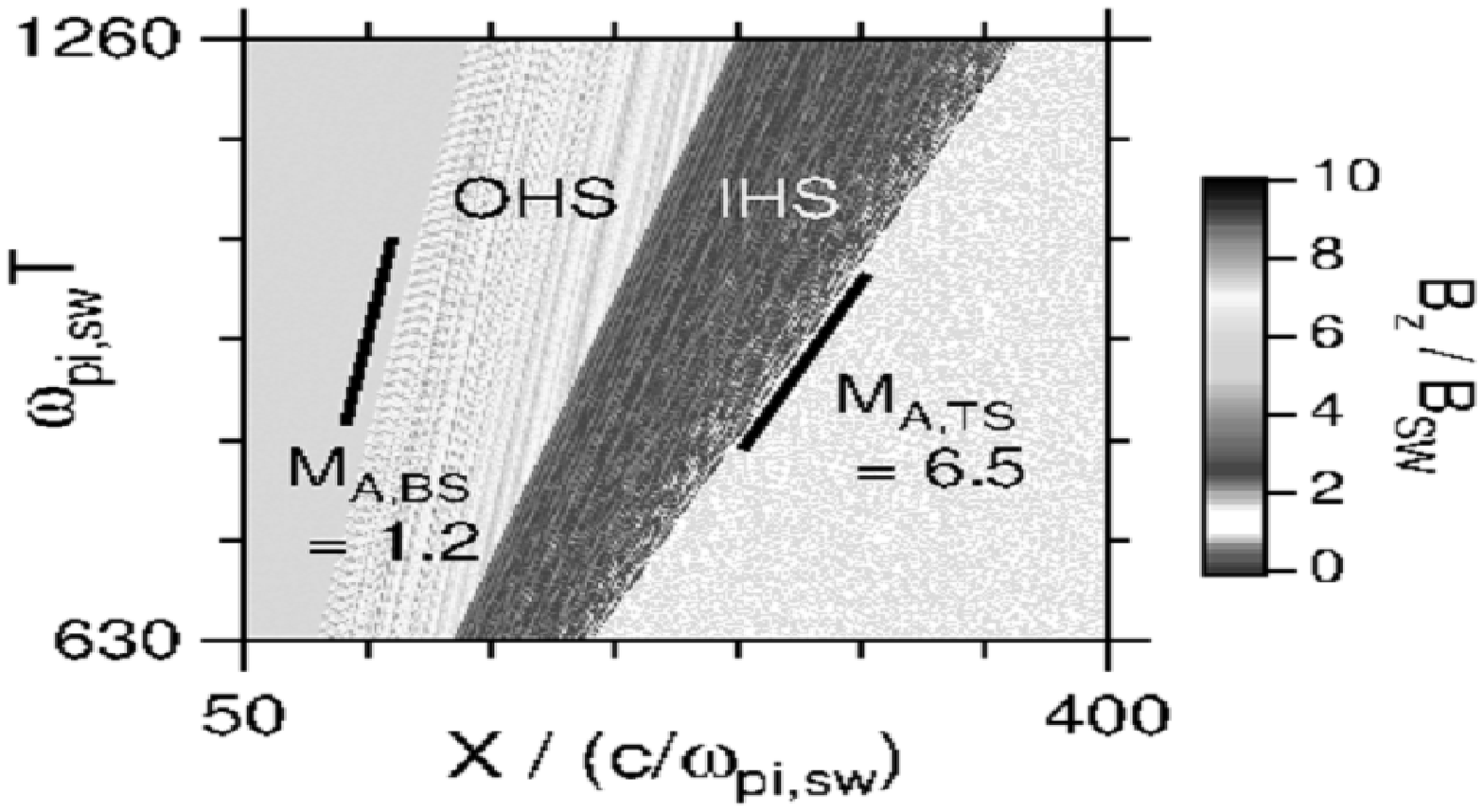}
%\plottwo{f1_color.eps}{f1.eps}
\caption{Spatio-temporal evolution of the magnetic field $B_z$ component for Run 1.
\label{fig_st1}}
\end{figure}

Fig.\ref{fig_st1} shows the spatio-temporal evolution of the magnetic field  $B_z$ component of 
Run 1. It is clear that there are four distinct regions separated by three discontinuities. 
The three discontinuities are, from the right, a forward shock, a contact discontinuity, 
and a reverse shock, in the context of the shock tube problem. They correspond to the 
termination shock (TS) for the SW, the heliopause (HP), and the bow shock (BS) in IS space, 
respectively. The four distinct regions from the right correspond to the unshocked SW, 
the shocked SW/inner heliosheath (IHS), the shocked IS plasma/outer heliosheath (OHS), 
and the unshocked IS plasma, respectively. From the slope associated with a discontinuity, 
the Alfv\'{e}n Mach number of the TS is estimated as $M_{A,TS} \approx 6.5$ and that 
of the BS as $M_{A,BS} \approx 1.2$. If one assumes time stationarity of each shock, 
these values and the values of upstream plasma beta and shock angle with specific 
heat ratio $\gamma = 2$ yields a theoretical compression ratio of the two shocks 
as 2.8 and 1.2, respectively. 

\begin{figure}
\epsscale{0.5}
\plotone{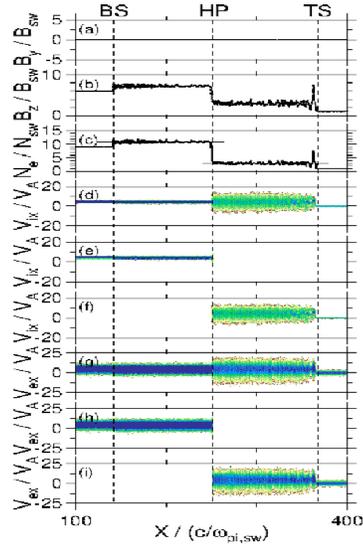}
%\plottwo{f1_color.eps}{f1.eps}
\caption{Field profiles and phase space densities at $\omega_{pi,sw}T=1260$ for Run 1. 
From the top, magnetic field (a) $B_y$, (b) $B_z$, (c) electron density $N_e$, 
phase space density of (d) all ions, (e) IS ions, (f) SW ions, (g) all electrons, 
(h) IS electrons, and (i) SW electrons, respectively, are plotted. The horizontal dashed 
lines in (c) denote the downstream densities of the two shocks inferred 
from the Rankine-Hugoniot relations.
\label{fig_prof1}}
\end{figure}

Fig.\ref{fig_prof1} denotes, from the top, profiles at $\omega_{pi,SW}T = 1260$ of the 
(a) magnetic field 
$B_y$ component, (b) $B_z$ component, (c) electron density $N_e$, (d) phase 
space density of all ions, (e) IS ions, (f) SW ions, (g) that of all electrons, 
(h) IS electrons, and (i) SW electrons, respectively. The three discontinuities 
are again clearly seen in all the panels except for panel (a) which is the incompressible 
magnetic field component that does not appear in the exactly perpendicular geometry 
case ($\Theta_{Bn,IS} = \Theta_{Bn,SW} = 90^{\circ}$). The horizontal dashed 
lines in panel (c) denote the values of the density inferred from the Rankine-Hugoniot 
relations as mentioned above. From panels (d)-(i), it is confirmed that the IS plasma 
and the SW plasma are clearly separated at the HP, although a small fraction of both 
ion species can be found roughly within a distance of a typical ion gyro radius. 
While in the oblique 
cases (Runs 2-5) some particles propagate across the HP along the magnetic field 
lines, the basic features above continue to hold.

It is notable that fluctuations in density as well as the magnetic field $B_z$ component 
are apparent in both the IHS and the OHS (see Fig.\ref{fig_st1} and Fig.\ref{fig_prof1}). 
The origin of the fluctuations will be discussed later.

%%%%%%%%%%
\subsection{Structure of Contact Discontinuity}
%%%%%%%%%%
%%%%%%%%%%%%%
\subsubsection{Spatial Scale}
%%%%%%%%%%%%%

\begin{figure}
\epsscale{0.5}
\plotone{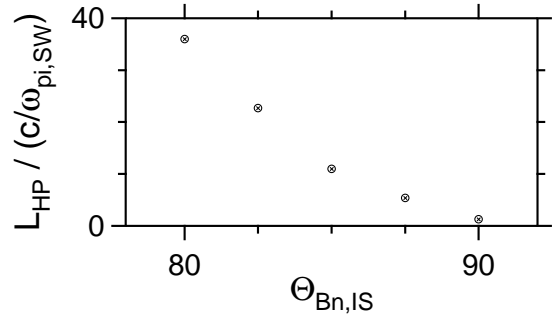}
%\plottwo{f1_color.eps}{f1.eps}
\caption{Spatial width of the contact discontinuity as a function of magnetic field 
obliquity $\Theta_{Bn, IS}$ with respect to IS plasma.
\label{fig_hpwidth}}
\end{figure}

\begin{figure}
\epsscale{0.5}
\plotone{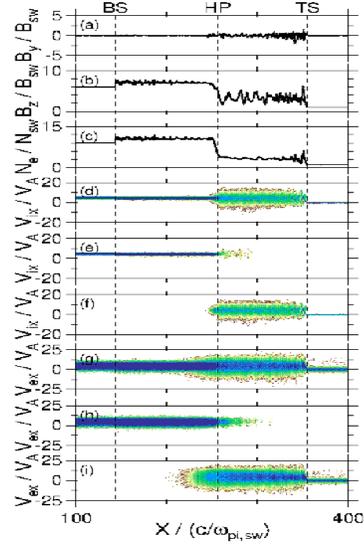}
%\plottwo{f1_color.eps}{f1.eps}
\caption{Field profiles and phase space densities at $\omega_{pi,sw}T=1260$ for Run 3 
(same format as Fig.\ref{fig_prof1}). 
\label{fig_prof3}}
\end{figure}

The dependence of the spatial scale of the contact discontinuity (HP) with $\Theta_{Bn}$ 
is examined here. Fig.\ref{fig_hpwidth} shows the widths of the HP, $L_{HP}$, at 
$\omega_{pi,sw}T = 1260$ for all the runs. The width is 
defined as the distance between the two points at which the density deviates 
from its averaged shock downstream value by $1N_{e,SW}$. 
In an oblique geometry ($\Theta_{Bn,IS} < 90^{\circ}$), particles can cross the 
HP as they move along a magnetic field line. This results in mixing of the IS 
plasma and the SW plasma near the HP, as seen in Fig.\ref{fig_prof3} for Run 3, 
and makes the width of the HP larger than that of the perpendicular case (Run 1). 
The mixing region expands as the magnetic field becomes more oblique. 

Note that when $\Theta_{Bn} < 90^{\circ}$, $B_z$ should be conserved across 
the HP if the HP is a contact discontinuity. However, this is not true in the 
runs 2-5 (e.g., Fig.\ref{fig_prof3}(b)). Indeed, we confirmed that the transverse 
momentum is not conserved across the HP in these runs. Therefore, the HP 
here is no longer a contact discontinuity formally. This may indicate 
that the system has not yet reached a steady state. However, we assume that 
the following features hold in the actual HP.

%%%%%%%%%%%%%
\subsubsection{Inner Structure}
%%%%%%%%%%%%%

\begin{figure}
\epsscale{0.5}
\plotone{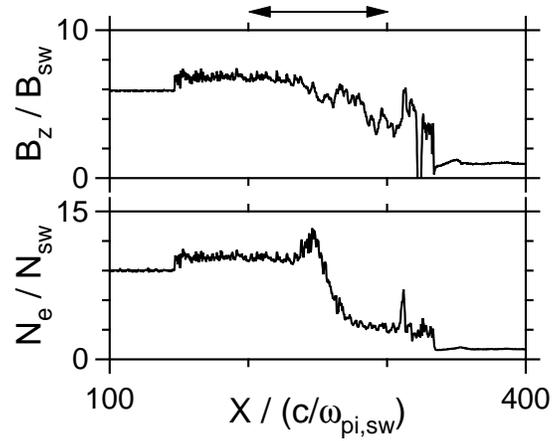}
%\plottwo{f1_color.eps}{f1.eps}
\caption{Spatial profiles of $B_z$ and $N_e$ for Run 5. 
\label{fig_BN80}}
\end{figure}

\begin{figure}
\epsscale{0.5}
\plotone{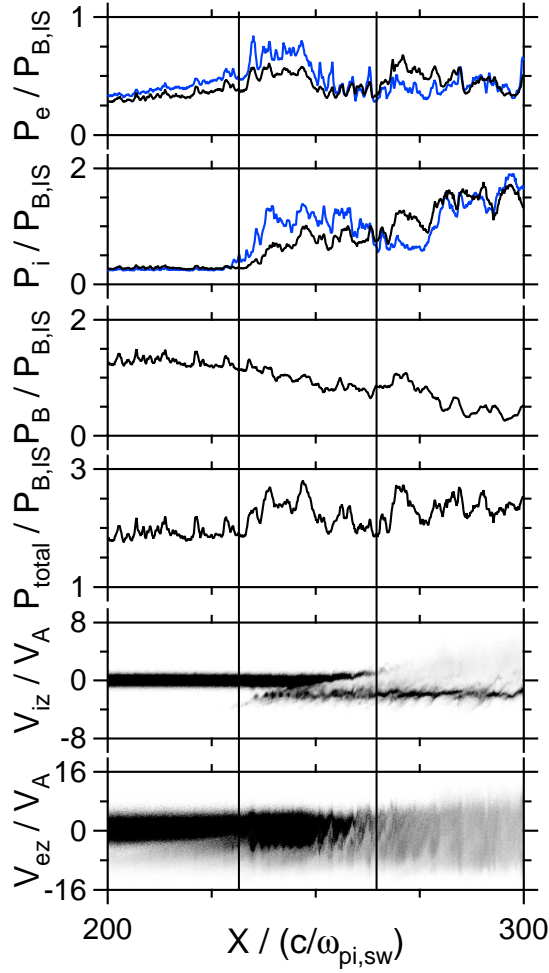}
%\plottwo{f1_color.eps}{f1.eps}
\caption{Expanded view of the structure near the HP for Run 5. From the top, 
electron pressure 
parallel (blue) and perpendicular (black) to the magnetic field, ion pressure parallel (blue) 
and perpendicular (black) to the magnetic field, magnetic pressure, total pressure, 
ion phase space density in $V_z-X$, and electron phase space density in $V_z-X$, 
respectively. The density hump in Fig.\ref{fig_BN80} 
corresponds to the region between the two vertical lines.
\label{fig_pressure80}}
\end{figure}

The deviation of the spatial profiles between the magnetic field and density in the HP 
becomes significant when $\Theta_{Bn}$ becomes smaller. In particular, a clear 
hump appears only in the density when $\Theta_{Bn,IS}=80^{\circ}$ (Run 5)  as shown 
in Fig.\ref{fig_BN80} and $\Theta_{Bn,IS}=82.5^{\circ}$ (Run 4: not shown). The detailed 
structure near the hump in the region indicated by the arrow in Fig.\ref{fig_BN80}, is 
illustrated in Fig.\ref{fig_pressure80}. In the density hump, between the two vertical lines, 
the parallel pressure of the ions (and electrons) dominates the total pressure. This 
region coincides with the region where SW ions and IS ions overlap 
in phase space, as shown in the fifth panel. These two ion populations interpenetrate 
as they move along the magnetic field. As a result, the local density is 
enhanced. Furthermore, the two populations are well separated in the $V_z-X$ 
phase space so that the local pressure is effectively well enhanced along the 
magnetic field direction. The electrons basically follow the ions and show a 
similar feature, while their effective parallel pressure is enhanced even 
deeper in the OHS where part of the hotter IHS electrons penetrate further. 
Note that this structure is a non-MHD effect.

%%%%%%%%%%%
\subsection{Fluctuations Downstream of Shocks}
%%%%%%%%%%%

The downstream region of the two shocks is turbulent in all the runs. This is clearly 
confirmed, for instance, in Figs.\ref{fig_st3} and \ref{fig_st5} showing the spatio-temporal 
evolution of $B_y, B_z$, and $N_e$ in Run 3 and Run 5, respectively. The IHS 
contains fluctuations of all three components, while the OHS contains 
fluctuations of only two components, $B_z$ and $N_e$. This is true for 
Run 2 to Run 5.

%%%%%%%%%%%%%
\subsubsection{Fluctuations in IHS}
%%%%%%%%%%%%%

\begin{figure}
\epsscale{0.5}
\plotone{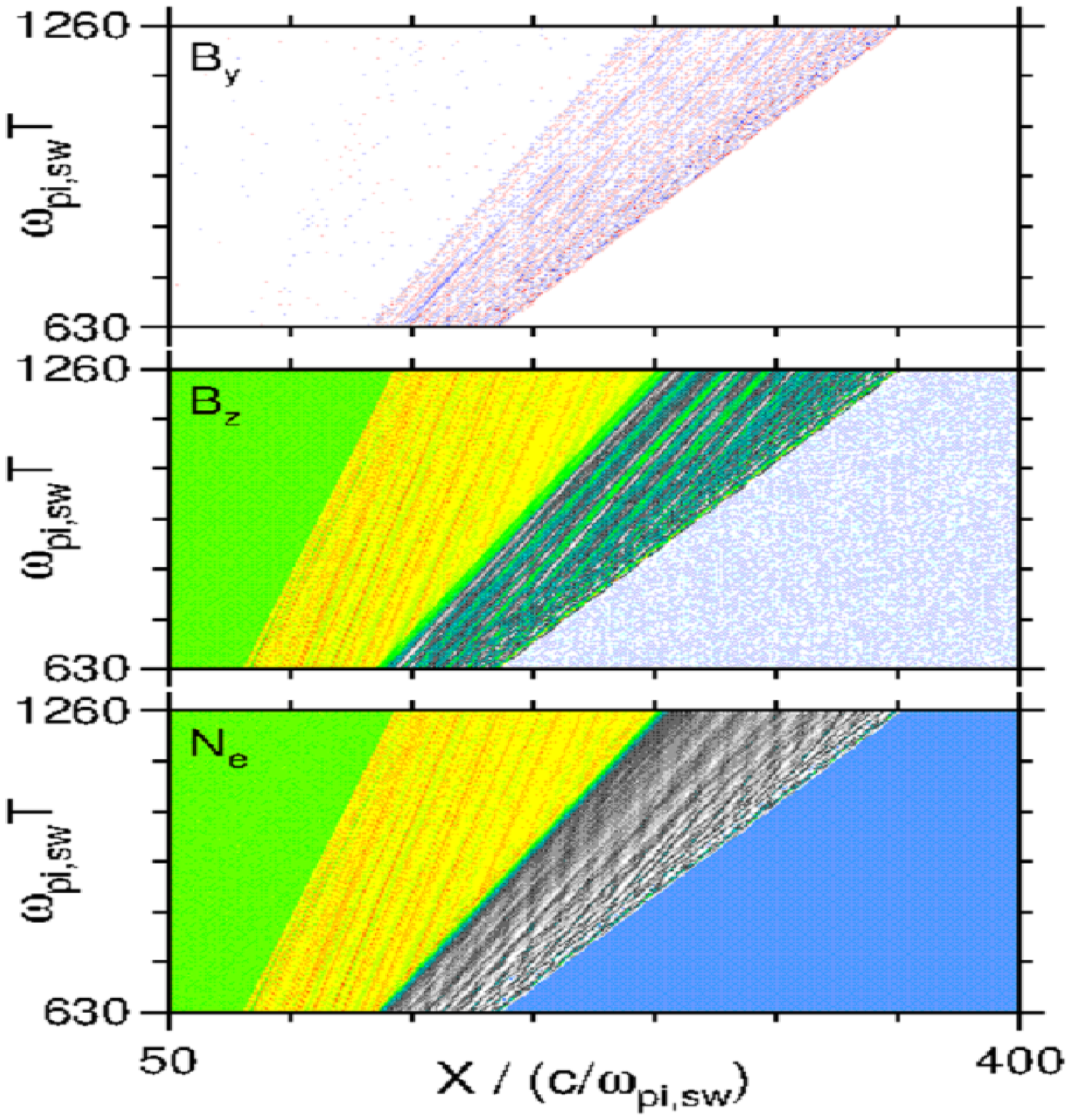}
%\plottwo{f1_color.eps}{f1.eps}
\caption{Spatio-temporal evolution of the incompressible ($B_y$) and compressible ($B_z$) 
magnetic fields, and electron density ($N_e$) for Run 3.
\label{fig_st3}}
\end{figure}

\begin{figure}
\epsscale{0.5}
\plotone{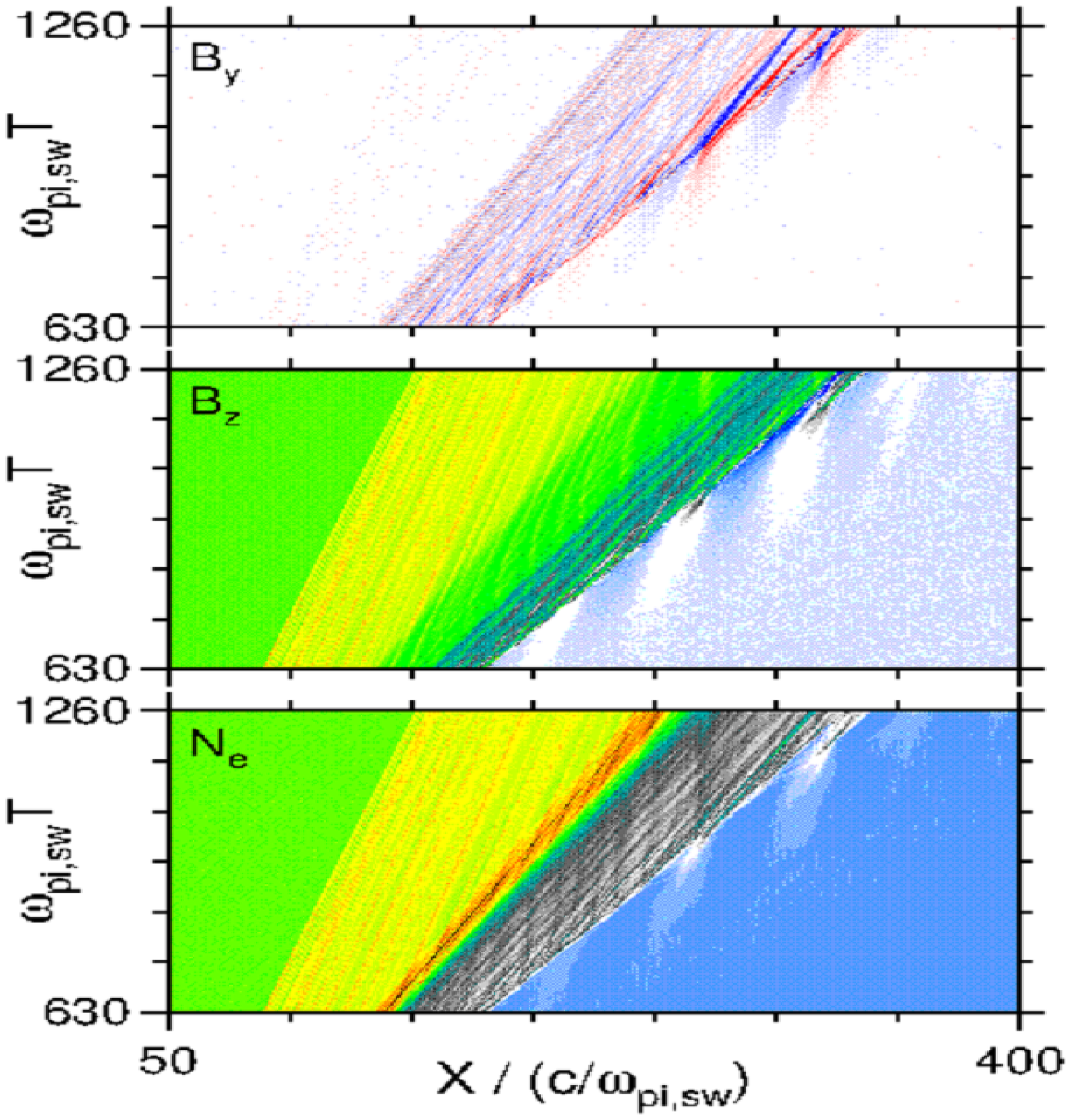}
%\plottwo{f1_color.eps}{f1.eps}
\caption{Spatio-temporal evolution of the incompressible ($B_y$) and compressible ($B_z$) 
magnetic fields, and electron density ($N_e$) for Run 5.
\label{fig_st5}}
\end{figure}

\begin{figure}
\epsscale{0.5}
\plotone{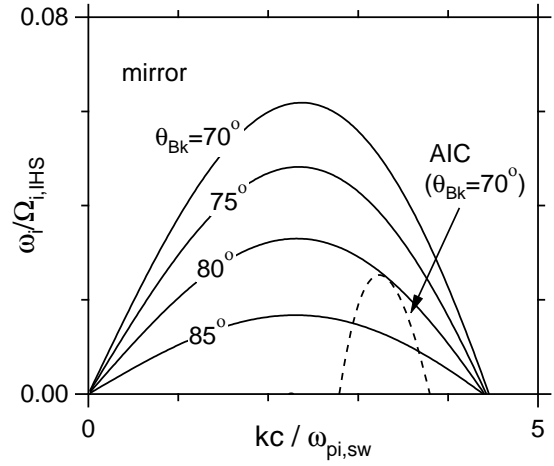}
%\plottwo{f1_color.eps}{f1.eps}
\caption{Linear growth rate of instabilities driven by ion temperature anisotropy in the IHS. 
The solid lines indicate the mirror instability for various wave propagation angles. The dashed 
line denotes the AIC instability. See the text for details.
\label{fig_AICgrowth}}
\end{figure}

One possible source of the fluctuations in the IHS is the temperature anisotropy. 
An ion temperature anisotropy may drive instabilities such as the Alfv\'{e}n ion cyclotron 
(AIC) instability or an electromagnetic ion cyclotron instability and mirror instability. 
In Run 1 the ion temperature perpendicular 
to the magnetic field is roughly 20 times higher than that parallel to the magnetic 
field in the IHS ($T_{i\perp, IHS} / T_{i\parallel, IHS} \sim 20$). Using this value and 
other parameters in the IHS estimated from the simulation, the growth 
rate of the instabilities are calculated numerically by solving the linear dispersion 
relation for a bi-Maxwellian hot plasma as shown in Fig.\ref{fig_AICgrowth}. Here, the ratio 
of local electron plasma frequency to cyclotron frequency is fixed as 
$\omega_{pe,IHS} / \Omega_{e,IHS} = 4$ for simplicity (We confirmed that the 
dependence on this ratio is rather small.). The solid lines denote the linear growth 
rate of the mirror instability for various wave propagation angles with respect to 
magnetic field, $\theta_{Bk}$. The values of $\theta_{Bk}$ correspond roughly to 
those in the IHS for Run 2 ($\theta_{Bk}=85^{\circ}$) and to Run 5 
($\theta_{Bk}=70^{\circ}$).
The dashed line indicates the linear growth rate of the AIC instability at 
$\theta_{Bk}=70^{\circ}$ (Run 5). The growth rate of this instability is too 
small or even negative when $\theta_{Bk} \ge 75^{\circ}$. By comparing this 
linear analysis with Figs.\ref{fig_st3} and \ref{fig_st5}, we conclude that the apparent 
wavy structures that are phase standing with respect to the HP 
seen in $B_z$ of Figs.\ref{fig_st3} and \ref{fig_st5} are due to the mirror instability.

Self-reformation of the surface of TS is another source of IHS fluctuations 
in this particular simulation.  Since the TS is a fast 
mode shock, the reformation leads to in-phase oscillations between $B_z$ and 
$N_e (N_i)$. This type of oscillation is present in all the runs (e.g., 
Fig.\ref{fig_st5}) and they have amplitudes as large as the waves generated by 
the linear instabilities discussed above. Such large amplitude waves 
can further be sources of other waves through nonlinear processes, like 
wave-wave couplings, which are not discussed in this paper. As a result, the IHS 
in each run contains a variety of wave modes, fast, slow, intermediate modes, 
and mirror modes, propagating in both positive and negative directions.

Besides the above, there are undoubtedly other causes of fluctuations in the 
actual IHS. Disturbances in the solar wind can be a dominant source of the 
fluctuations in IHS \citep{donohue93, story97, zank03, washimi11}. 
Turbulent magnetic 
reconnection occurring in the IHS is another possible source. Regardless 
of the source of the fluctuations, we discuss the possibility that the 
fluctuations in the IHS are transmitted through the HP into the OHS.

%%%%%%%%
\subsubsection{Fluctuations in the OHS}
%%%%%%%%

In Figs.\ref{fig_st3} and \ref{fig_st5} fluctuations in the OHS are visible in $B_z$ and 
$N_e$, while $B_y$ exhibits no signal in the same region. Here, $B_y$ indicates the 
incompressible component of the magnetic fluctuations. Clearly, the 
incompressible fluctuations produced in the IHS are confined in the same 
region and are not transmitted to the OHS. On the other hand, the OHS 
fluctuations in $B_z$ and $N_e$, which are compressible components, 
clearly propagate away from the HP.

\begin{figure}
\epsscale{0.5}
\plotone{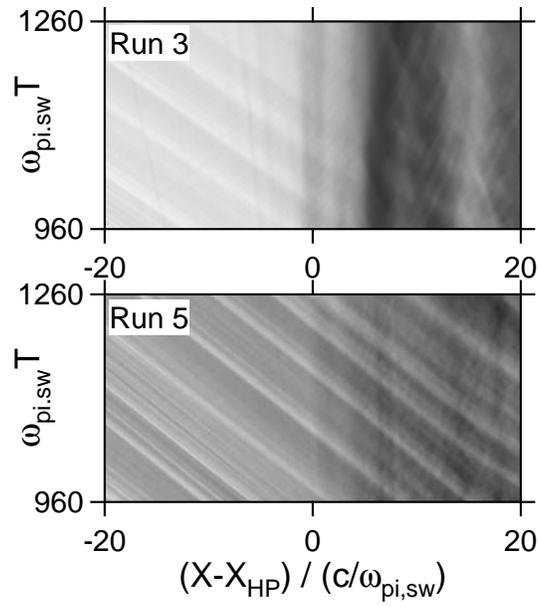}
%\plottwo{f1_color.eps}{f1.eps}
\caption{Details of wave propagation near the HP for Run 3 (upper panel) and Run 5 (lower 
panel). In both panels the gray scale denotes the magnitude of $B_z$. The horizontal axis 
is the relative distance from the HP and the vertical axis is time.
\label{fig_expHP}}
\end{figure}

Fig.\ref{fig_expHP} is an expansion of the middle panels ($B_z$) of 
Figs.\ref{fig_st3} (Run 3) and \ref{fig_st5} (Run 5). 
The horizontal axis denotes the relative distance from the HP. It is clear 
that the wave peaks in the OHS ($X < X_{HP}$) are continuously linked with 
those in the IHS ($X > X_{HP}$), indicating that the waves originated in 
the IHS, although other waves are also present in the IHS. 

\begin{figure}
\epsscale{0.5}
\plotone{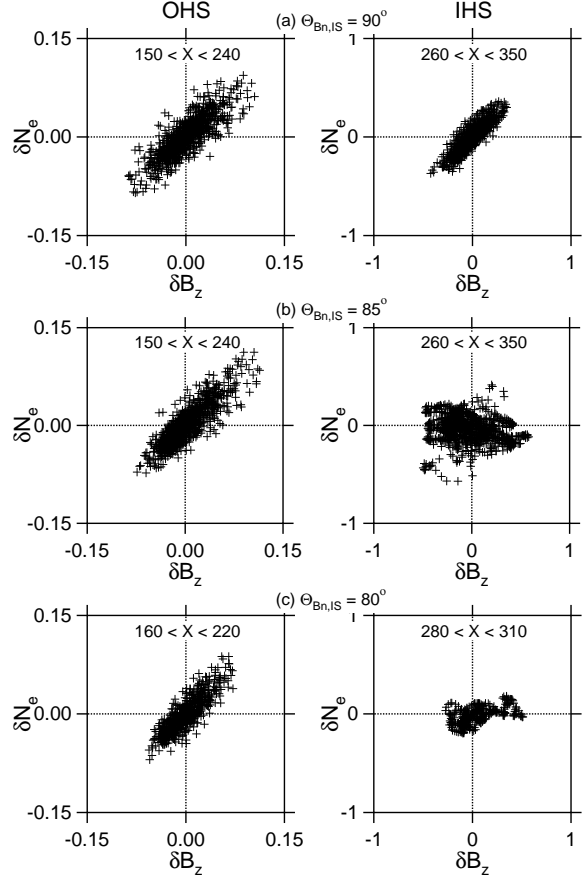}
%\plottwo{f1_color.eps}{f1.eps}
\caption{Scatter plots between density ($\delta N_e$) and compressible magnetic 
($\delta B_z$) fluctuations in the OHS (left panels) and the IHS (right panels) for 
(a) Run 1, (b) Run 3, and (c) Run 5. 
\label{fig_scat}}
\end{figure}

The correlation between the magnetic and density fluctuations in the IHS 
and OHS for different runs is shown in Fig.\ref{fig_scat} as a scatter plot of 
density versus magnetic fluctuations. It shows a clear positive correlation 
between them in the OHS as seen in the left panels, even when the two 
components do not have clear positive correlations in the IHS (right panels).
This indicates that among the compressible waves generated in the IHS, 
only the fast mode waves can pass the HP into the OHS in the present simulations.

%%%%%%%%%%%
\subsubsection{Effect of the HP}
%%%%%%%%%%%

\begin{figure}
\epsscale{0.5}
\plotone{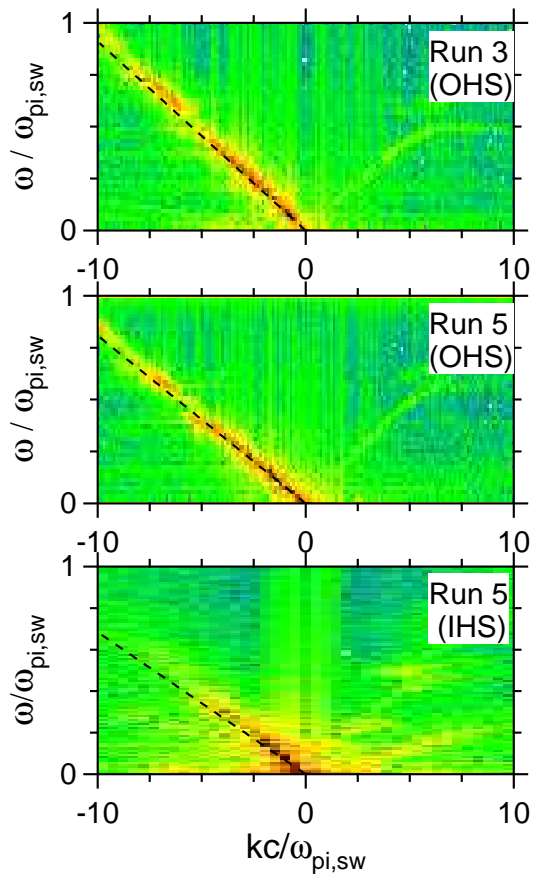}
%\plottwo{f1_color.eps}{f1.eps}
\caption{$\omega-k$ spectra of $B_z$ for Run 3 (upper panel) and Run 5 
(middle and bottom panels). The corresponding time interval is 
$941 \le \omega_{pi,sw}T \le 1265$. The spatial region is 
$-56.87 \le (X - X_{HP}) / (c / \omega_{pi,sw}) \le -5.06$ for the upper and 
the middle panels, while $5.06 \le (X - X_{HP}) / (c / \omega_{pi,sw}) \le 56.87$ 
for the bottom panel.
\label{fig_wk}}
\end{figure}

Fig.\ref{fig_wk} denotes the $\omega-k$ spectra of $B_z$ in the OHS in 
Run 3 (upper panel) and Run 5 (middle panel), and in the IHS in Run 5 (bottom 
panel) in the frame moving with the HP. In the OHS the dominant peak appears 
in the second quadrant. This is consistent with the 
OHS fluctuations propagating away from the HP in Figs.\ref{fig_st3} and \ref{fig_st5}. 
The black dashed lines in both (upper and middle) panels show the local Alfv\'{e}n 
velocity, which is close to the phase velocity of local fast mode waves, 
since the local plasma in the OHS has a low plasma beta. In the IHS the 
dominant peak again appears in the second quadrant slightly above the 
black dashed line indicating the local Alfv\'{e}n velocity, while other wave 
modes can also be recognized in the first and second quadrants. The difference 
between the bottom two panels again indicates that only the fast mode waves 
existing in the IHS can cross the HP into the OHS (VLISM).

\begin{figure}
\epsscale{0.5}
\plotone{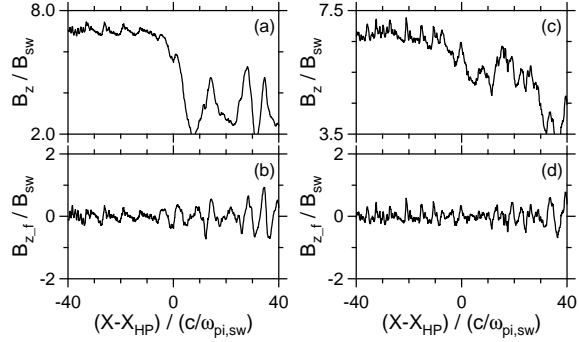}
%\plottwo{f1_color.eps}{f1.eps}
\caption{Expanded view of $B_z$ near the HP for (a) Run 3 and (c) Run 5. 
The panels (b) and (d) are obtained from (a) and (c), respectively, by applying 
a high pass filter where fluctuations of wavelength $4.3 c/\omega_{pi,sw}$ 
or larger are filtered out.
\label{fig_Bzfilter}}
\end{figure}

This scenario indeed is theoretically and numerically supported. \cite{zank17} show 
that the conditions at the heliospheric boundary region allow only fast mode waves 
in the IHS to cross the HP by considering the so-called Snell's law. It is also 
shown by \cite{washimi07, washimi11, washimi17} using their MHD simulation that large 
amplitude fast mode disturbances in the solar wind are partially transmit throught 
the HP. However, 
the simulation result here appears to show a discrepancy from the theory based 
on the Snell's law. 
Although in theory, the wave frequency should not change across the HP, 
the frequency of the fast mode waves increases after crossing the HP. 
One possible explanation for this is wave steepening. Fig.\ref{fig_Bzfilter} (a) 
and (c) denote the expanded view of $B_z$ near the HP for Run 3 and Run 5, 
respectively. The waveforms in $X < X_{HP}$ is clearly steepened, although 
no steepened feature is seen in $X > X_{HP}$. This is confirmed more clearly 
in (b) and (d) where the fluctuations having wavelength of $4.3 c/\omega_{pi,sw}$ 
or longer are filtered out from (a) and (c), respectively. The fast mode waves 
generated in the IHS are steepened when crossing the HP. This may result in 
broadening the wave spectrum in the OHS.

%First, it is clear that the intermediate waves, i.e., incompressible fluctuations, cannot 
%penetrate the HP, since the Alfv\'{e}n speed in the OHS is always different from 
%that in the IHS so that eq.(\ref{snell}) is never satisfied. For fast and slow waves, 
%i.e., compressible fluctuations, the left and the right hand sides in eq.(\ref{snell}) 
%are plotted as a function of the wave propagation angle in Fig.\ref{fig09}. The solid 
%lines denote the left hand side and the dashed lines indicate the right hand side of 
%eq.(\ref{snell}), respectively. The red and the 
%green lines show fast and slow waves. Here, we have assumed that the plasma 
%beta in the IHS is $\beta_{IHS}=4$ and that in the OHS is $\beta_{OHS}=0.3$, and 
%the OHS to IHS Alfv\'{e}n velocity ratio is $V_{A,OHS}/V_{A,IHS}=1.3$, respectively. 
%These are typical values in the series of simulations. Furthermore, the specific 
%heat ratio is assumed to be $\gamma = 5/3$. As in \cite{zank17}, in terms of the 
%slow waves (green lines), the left hand side (solid line) is never equal to the right 
%hand side (dashed line), whereas eq.(\ref{snell}) can be satisfied in terms of 
%the fast waves (red lines) for a wide range of $\theta_f$. This indicates that only the 
%fast waves generated in the IHS can pass the HP and be transmitted into the OHS. 

%%%%%%%%%%%
\section{Summary and Discussions}
%%%%%%%%%%%

In this study one-dimensional PIC simulations were performed to investigate 
the so-called shock tube problem in a collisionless plasma in which a relatively 
tenuous and weakly magnetized plasma is continuously pushed by a relatively 
dense and strongly magnetized plasma having supersonic relative velocity. 
The structure of the boundary region includes three discontinuities, 
forward and reverse shocks and a contact discontinuity. 
The spatial width of the contact discontinuity increases as $\Theta_{Bn}$ 
decreases. The spatial profiles of the magnetic field and plasma density, or 
pressure, of the contact discontinuity change significantly when 
$\Theta_{Bn,IS}$ deviates from 
$90^{\circ}$. This is a non-MHD or kinetic effect of the local plasma that allows 
charged particles to stream across the contact discontinuity. Many 
kinds of magnetic and density fluctuations are observed in the region between 
the two shocks. Among them, the fluctuations between the reverse shock 
and the contact discontinuity (OHS) are fast magnetosonic modes and propagate 
away from the contact discontinuity. Their origin is on the IHS 
side of the contact discontinuity. 

If we regard the above boundary structures as mimicing the heliospheric 
boundary, the forward and the reverse shocks correspond to the solar wind 
termination shock and the bow shock, and the contact discontinuity to the 
heliopause, respectively. In the context of heliospheric physics 
the TS is a reverse shock and the BS, if present, is a forward shock.

\begin{figure}
\epsscale{0.5}
\plotone{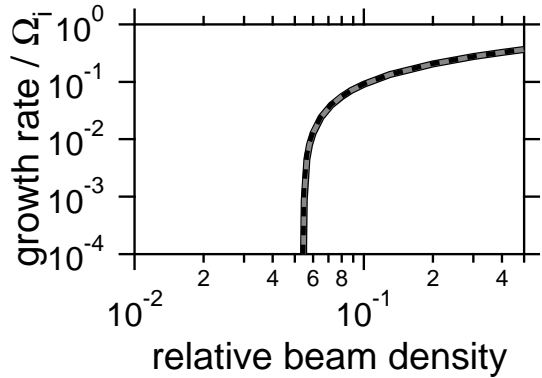}
%\plottwo{f1_color.eps}{f1.eps}
\caption{Linear growth rate of the right hand resonant ion beam instability 
that may occur in the density hump. For further details, see the text.
\label{fig_RIgrowth}}
\end{figure}

A difference in the profiles between the magnetic field and cosmic ray count 
rate at the heliopause is observed by Voyager 1 \citep{burlaga14}. This 
indicates that the observed cosmic rays are not tied to the local magnetic 
field within the scale of the changing magnetic field. This is consistent with 
our simulation result that even the background 
plasma does not satisfy the frozen-in condition locally in the heliopause. From 
the simulation, the plasma density or pressure is more enhanced than the magnetic 
field in the heliopause. This discrepancy between the plasma and the magnetic 
field profiles may be sustained unless the separation between the SW plasma 
and the IS plasma in the local phase space is relaxed. Such relaxation may 
occur through either collisionless or collisional processes. \cite{mostafavi18} 
showed that collisional mean free paths of various kinds (p-p, e-p) in the VLISM are 
typically of the order of 0.1AU. On the other hand, collisionless instabilities 
may occur through ion-ion two stream interactions. According to 
\cite{richardson18}, plasma parameters observed by Voyager 2 in the IHS 
after 2018 are typically $N \sim 2 \times 10^{-3} cm^{-3}$, $T \sim 5 eV$, 
and $|V| \sim 10^2 km/s$, respectively. Here, $N$ is the density, $T$ the 
thermal proton temperature, and $|V|$ denotes the bulk flow speed. The typical 
magnetic field strength also observed by Voyager 2 in the IHS in 2015 is of the 
order of $0.1 nT$ \citep{burlaga18b}, which is consistent with Voyager 1 
observations in the IHS \citep{burlaga14b}. From these values, the local Alfv\'{e}n 
velocity is estimated as $V_A \sim 50 km/s$. In the OHS or VLISM, the magnetic field and thermal plasma 
density increase, although an exact value of the latter is not yet well known. 
If the two plasmas interpenetrate in the heliopause transition region as seen 
in the simulation, the right hand resonant instability may possibly be generated. 
By way of illustration, we calculated the linear growth rate of the 
instability as a function of relative beam density, which is defined as the SW 
proton density divided by the IS proton density, for the case that the relative 
bulk velocity of the two warm proton beams is $2 V_A$ (Fig.\ref{fig_RIgrowth}). 
The temperature of the SW protons is assumed to be 5eV, while that of the IS 
protons is taken to be 1eV (black solid line) and 25eV (gray solid line although almost hidden 
by the black solid line). We also assume that there 
are two non-thermal populations. One is PUIs whose relative density is 25\% and 
temperature is $100eV$. Another non-thermal population is assumed to be present 
and to have a much 
lower relative density, 1\%, and much higher temperature such as $10keV$ 
(black and gray solid lines) and $1MeV$ (black dashed line). This component 
might be related to anomalous or galactic cosmic rays for example. We find that the 
resultant growth rates of the instability do not much depend on the above 
temperature variations and that the growth rate becomes of the order of 
$10^{-2} \Omega_i$ or larger when the relative beam density reaches roughly 
6\%. Here, $\Omega_i$ denotes proton cyclotron frequency. 
The local magnetic field is several times larger than $0.1nT$ \citep{burlaga14} 
so that the corresponding time scale of the wave growth is a few thousand $sec$. 
During this time, the plasma, having a typical bulk flow speed of $100 km/s$, travels 
roughly $\sim 10^5 km$ ($10^{-2} \sim 10^{ -3} AU$) which is much less 
than the collisional mean free path. As illustrated in Fig.\ref{fig_RIgrowth}, the growth 
rate of the instability is even higher for larger relative beam densities so that the 
above effective mean free path may have been overestimated. Another known 
non-MHD structure in boundary regions is the plasma depletion layer (PDL), 
which is observed in front of planetary magnetopauses and thought to exist 
beyond the HP \citep{fuselier13, cairns17}. The spatial scale of the PDL along 
the trajectory of Voyager 1 is estimated as 2.6AU from the Plasma Wave Subsystem 
as well as the Magnetometer instrument data on board Voyager 1. Therefore, 
the density hump discussed here has a quite different spatial scale from the PDL. 

\cite{burlaga18} showed that magnetic turbulence in the VLISM observed 
by Voyager 1 during an interval from 2013.3593 to 2014.6373 is more or less 
compressible. During that time Voyager 1 was rather close to the 
heliopause. Possible sources of the compressible fluctuations are discussed 
by \cite{burlaga18}. A bow shock (or a bow wave) is one possibility. However, the 
bow shock is expected to be very weak, even if it exists \citep{mccomas12, zank13}. 
Although our simulation 
reproduced a weak reverse shock, dominant fluctuations do not propagate from 
the reverse shock but do propagate toward it. Hence, a bow shock is unlikely 
to be a source of the observed turbulence as long as it is very weak. A second 
possibility raised by \cite{burlaga18} is that the turbulence is generated in the 
OHS or VLISM. Furthermore, they mention that the heliopause itself can also 
be a source. These possibilities are not negated from our simulations. The 
fourth possibility is that the fluctuations might originate in the IHS and pass 
through the heliopause into the OHS or VLISM, which is suggested by 
\cite{burlaga15} and theoretically demonstrated by \cite{zank17}. Our simulation 
here supports this scenario. The heliopause works as a filter at which incompressible 
fluctuations and compressible slow mode fluctuations present in the IHS are 
filtered out, and only the compressible fast mode fluctuations can pass through 
it. This is what was expected theoretically by \cite{zank17}. In our simulation 
it is also shown that the wave spectrum in the OHS or VLISM is broadened. 
A possible explanation for this is the wave steepening, i.e., the fast mode fluctuations 
generated in the IHS are steepened when passing the HP. Another possible 
explanation may be based on the inhomogeneous flow velocity around the HP. 
When the mean flow velocity changes, which is actually confirmed in Run 5 
($\Theta_{Bn, IS} = 80^{\circ}$), the wave frequency can change as a result 
of the conservation of wave action \citep{bretherton68, zank87}.

The simulation here is one-dimensional. All higher dimensional structures 
and phenomena in the boundary region are ignored. Also only the main population of 
the solar wind has been taken into account. 
%{\bf 
Further, some parameters used in the simulations are not realistic. 
%}
We comment briefly on these points below. 
On the kinetic scales of the 
problem, the scale of curvature of a heliopause located at 124 AU or so is 
essentially zero i.e., as far as the kinetic scale waves are concerned, the 
heliopause is a planar structure. Thus we neglect 
the fact that the streamline of the plasma near the heliopause gradually bends 
along its surface. The relative tangential flow speeds across the heliopause 
are not especially large and would result primarily in a Doppler-shift in the 
frequency without significantly modifying the physics of the problem (we are 
not considering a Kelvin-Helmholtz instability - see for example the discussion 
in \cite{avinash14} related to this point). Moreover, some important 
effects in the heliospheric boundary region, like charge exchange and pickup ion 
dynamics, are also not taken into account. As discussed above, charge-exchange 
can be neglected in the simulation. However, we believe that the 
presence of distinct non-thermal populations do not significantly alter the 
essential wave properties observed in the current simulations. The 
shock jump conditions determine the correct total thermal and 
kinetic energy in the inner heliosheath, regardless of whether the 
non-thermal populations are included explicitly or not. Nonetheless, it is fair to say that 
the distribution is not a thermal equilibrium Maxwellian distribution. Immediately downstream of the termination shock, both the thermal ions and the pickup 
ions exhibit temperature anisotropies \citep{matsukiyo07}. \cite{zank10} showed 
that the inner heliosheath distribution can be approximated rather well 
by an isotropic kappa distribution. The distribution function in the inner 
heliosheath that we compute does depart from a simple Maxwellian 
because of dissipative processes (ion reflection) at the heliospheric 
termination shock. In this regard, it probably captures somewhat reasonably 
the more complicated distribution function expected in the inner heliosheath 
(certainly better than can be expected from a 3D MHD-neutral H model). 
%{\bf 
Furthermore, some parameters such as the mass ratio and frequency ratio ($\omega_{pe,SW} / 
\Omega_{e,SW}$) are far from the realistic values. It is known that changing these 
parameters may result in a change of wave generation mechanisms. For example, 
a number of instabilities excited in the transition region of a collisionless shock show 
different parameter dependences \citep{wu84}. For this reason, we have not 
focused on the generation mechanism for waves seen in the 
region downstream of the termination shock in this paper. However, regardless of their generation 
mechanisms, we believe that the wave propagation properties across the HP discussed 
in this paper are correct at least qualitatively. 
%}

\acknowledgments

%We should like to thank Hiroki~Akamatsu and Susumu~Inoue 
%for useful comments and discussions. 
%This work was supported by Grant-in-Aid for Scientific 
%Research on Innovative Areas 21200050, Grant-in-Aid for 
%Scientific Research on Priority Areas 19047004 (R.~Y.), 
%and Grant-in-Aid for Young Scientists (B) 21740184 (R.~Y.) 
%and 22740323 (S.~M.).

This work was supported by Grant-in-Aid for Scientific Research (C) 
No.19K03953 from JSPS (S.~M). G.P.Z. acknowledges partial support 
by an IBEX subaward SUB 0000167/NASA 80NSSC18k0237 and 
a NSF EPSCoR RII-Track-1 Cooperative Agreement OIA-1655280.
%Scientists (B) No.22740323 (S.~M.), Grant-in-Aid from the 
%Ministry of Education, Culture, Sports, Science, and 
%Technology (MEXT) of Japan, No.~21740184, No.~19047004 (R.~Y.), 
%and Grant-in-Aid for Scientific Research on Innovative Areas 
%21200050. 

%%
%%bibliography
%%

%\clearpage

%figure[fig01]
%
%figure[fig02]
%
%figure[fig03]
%
%figure[fig04]
%
%figure[fig05]
%
%figure[fig06]
%
%figure[fig07]
%
%figure[fig08]
%
%figure[fig09]
%
%figure[fig10]
%
%figure[fig11]
%
%figure[fig12]
%
%figure[fig13]
%
%%figure[fig09bw]
%\begin{figure}
%\epsscale{0.5}
%\plotone{fig09.eps}
%%\plottwo{f1_color.eps}{f1.eps}
%\caption{The left and the right hand sides of eq.(\ref{snell}) as a function of 
%wave propagation angle. The red and green lines denote fast and slow 
%modes and the solid and the dashed lines show the left and the right hand 
%sides of eq.(\ref{snell}), respectively.
%\label{fig09}}
%\end{figure}
%
%figure[fig14]

\end{document}